\documentclass[a4paper]{spie} 
\usepackage[]{graphicx}
\usepackage{ifthen}

\usepackage{xspace}
\usepackage{array}
\usepackage{changes}
\usepackage{amsmath,amsfonts,amssymb}
\usepackage[colorlinks=true, allcolors=blue]{hyperref}
\usepackage{color}
\usepackage{makecell}
\newcommand{\verify}[1]{   \ifthenelse{\boolean{accept}}{#1}{\textcolor{red}{#1 (to be verified!)}}               }

\title{Evaluation of the performance requirements for a constraint system applied to thin shell active mirrors in space applications: an analysis of the loads to be counteracted during the operative lifetime}

\title{Requirements and conceptual design for a constraint system applied to thin-shell active mirrors}

\title{Lateral constraint for active mirrors with thin glass shell: analysis of the requirements and conceptual design.}

\title{Lateral constraint for thin glass shell: analysis of the requirements and conceptual design for a segmented active mirror. }

\author[a]{Marcello Agostino Scalera}
\author[b]{Runa Briguglio}
\author[b]{Ciro Del Vecchio}
\author[b]{Marco Xompero}
\author[a]{Marco Riva}

\affil[a]{INAF Osservatorio Astrofisico Merate, Via E. Bianchi 46, 23807 Merate (LC),Italy }
\affil[b]{INAF Osservatorio Astrofisico Arcetri, L. E. Fermi 5, 50125 Firenze (FI), Italy }

\authorinfo{Further author information:\\Marcello Agostino Scalera: E-mail: marcello.scalera@inaf.it, Telephone: +39 348 5204516\\}

\begin{document} 

  \maketitle

  \begin{abstract}
The latest high-performance telescopes for deep space observation employ very large primary mirrors that are made of smaller segments, like the JWST which employs monolithic beryllium hexagonal segments. A very promising development stage of these systems is to make them active and to operate on their reflective surfaces to change their shape and compensate for aberrations as well as to perform a very precise alignment. This is possible by employing a reference body that stores actuators to modify the shape of the shell, like in the SPLATT project where voice coil actuators are used. However, the lack of physical contact between the main body and shell places – along with the many advantages related to the physical decoupling of the two bodies - some concerns related to the retaining of the shell under all the possible acceleration conditions affecting the system during the mission lifetime. This paper aims to study the acceleration environment affecting the spacecraft during its lifetime and to use it as a baseline for operational requirements of a retaining system for the shells. Any solution is selected in this paper to leave complete freedom for the development of a constraining system, just some are qualitatively discussed. 

   \end{abstract}


\section{Introduction}
In the LATT project, an ESA-funded TRP concluded in 2015, the concept of an active primary mirror (or segment) for a space telescope has been investigated. The team manufactured a 40 cm diameter demonstrator, called OBB (Optical Breadboard), actively shaped by 19 actuators. The technology is that used for large format adaptive secondary mirrors currently deployed at the Large Binocular Telescope\cite{10.1117/12.858229} (Arizona) and Very Large Telescope\cite{2014SPIE.9148E..45B} (Chile). The concept is based on the very favourable mix (as demonstrated on those adaptive optics systems on ground) of voice coil actuators (VCM) and thin glass shell (TS). The TS is a 1.5 to 2 mm thick Zerodur meniscus with magnets bonded on its back; the VCM are encapsulated in a rigid Reference Body (RB) and the TS is shaped thanks to the coupling VCM - magnets. In add, the VCM are controlled in a local close loop fed by co-located position sensor to drive the current in the coil and keep the wanted position, at a frequency much faster than the optical close loop with the wavefront sensor (WFS).\\
Such technology has been adopted for the Large 
The core of such technology (which is also adopted for the ELT M4\cite{2016SPIE.9909E..7YB} adaptive mirror and for the GMT adaptive M2) is then the TS, whose manufacturing process is now well mastered both in the USA and in Europe. For the M4 mirror, 8 TS have been  manufactured so far to realize the 2.5 m diameter optical surface, 2 mm thickness, and the typical manufacturing residual is lower than 20 nm RMS WF.\\
A TS is an extremely fragile piece of optics and its use on a space telescope is considered critical. One of the outcomes of the LATT project was the development and test of a constraint mechanism, based on an electrostatic force applied to the TS, to let it survive to the launch accelerations. The test was successful, yet the question "would you adopt a TS based active mirror on a space telescope?" is open. The answer could be positive when we focus on the global-level benefits of such technology, for instance the very low areal density: the OBB has an areal density lower than 16 kg/m2, including the support, actuators and position sensors. In add, VCM are contactless actuators: the mechanical gap between the RB (or the payload) and the optical surface would a natural insulator for mechanical vibrations.\\
Additional questions arise. Contactless actuation means that the TS are \emph{formation flying} in front of the telescope, is there the risk of loosing them? Do we need a retaining system to compensate such risk? Which are the functional requirements of such system to satisfy a safety need while preserving a low areal density and a mechanical insulation of the optical surfaces? Addressing in the correct view such questions could open the way to the use of very lightweight, large stroke active mirrors with global benefit for the mission.

Referring to the prototype of a thin shell actuated by voice coil actuators used in the SPLATT project, this paper studies the loads that affect the system during a JWST-like space mission, highlighting the most critical events like launch, orbital maneuvers during the orbital transfer, attitude maneuvers and orbital maintenance. The study is based mainly on the mission and operation profile data of the JWST and secondarily on those of the LUVOIR space telescope. Future JWST-like telescopes, such as Luvoir, are expected to take the largest advantage from the thin active shell technology used on the segmented primary mirror, but other applications can be foreseen for high-resolution telescopes for Earth observation or for solar system exploration, being the SPLATT prototype potential adaptable to any space payload. 
\newline
Following a system engineering conceptual approach, this paper aims to define the loads that the thin shell system shall sustain, placing some clear technical requirements to drive the development of the constraint system, without imposing any solution, but some qualitative hint at the end of the paper. The outcome of this study shall propose a guideline of the accelerations acting on the thin shell active systems, useful to design dedicated constraint systems or to strengthen the actuation technology. A clear distinction between one-off loads (launch, orbital transfer) and recurrent loads (attitude and orbital maintenance maneuvers) is highlighted since it may lead to different approaches and developments.

\section{One-off accelerations during the mission lifetime}
Launch accelerations are usually the highest ones experimented by a space system along its all complete lifetime and they must be supported only once. Being a one-off event, it is possible to think about non-permanent solutions that employ high energies that can be provided by the launcher and its batteries. A different situation is encountered while manoeuvring during the orbital transfer trajectory. Now, the accelerations are weaker than during launch but they are scheduled to happen in a way wider time frame and employ only the onboard resources without any external support. Once the orbital transfer is done, only the station keeping manoeuvres will affect the orbit of the spacecraft as recursive manoeuvres. Due to these profound operational differences, the study regarding the launch and manoeuvring phases are split and analyzed separately. 
    \subsection{Launch}
    The launch accelerations are summarised in the Ariane V launcher manual \cite{arianespace2016ariane}, where it is clearly stated that the longitudinal acceleration does not exceed 4.55g (value met at solid boosters' burnout) and the transverse accelerations do not exceed 0.25g. These values were the ones experienced by the JWST, but future spacecrafts will use different launchers like the new version of the SLS, already selected for the Luvoir launch. The available data for these new generation launchers are scarce and focused on specific mission profiles like lunar missions where they suggest lower accelerations that those generated by the Ariane V. Because of this, the data of the Ariane V have been used. 
    \newline
    The accelerations act differently on the shells according to the launch configuration inside the fairing. Referring to the launch configuration of the JWST and the expected one of Luvoir, the strongest longitudinal acceleration would be transferred to the shells as a force acting approximately on the plane of the shell surface, similarly to a shear force. This force tries to make the shell slide away from the reference body along the flight direction. The transverse accelerations act both on the shell surface and perpendicular to it. 
    \newline
    Another interesting analysis refers to the dynamical environment generated by the launcher. It is well described in the Ariane V user manual and can be used to evaluate the response of the shell to the vibration loads, but more importantly to this study, to compute the constraint forces to keep the shell in position.  
    \subsection{Orbital manoeuvres during interplanetary transfer}
\begin{figure}[h]
        \centering
        \includegraphics[scale = 0.5] {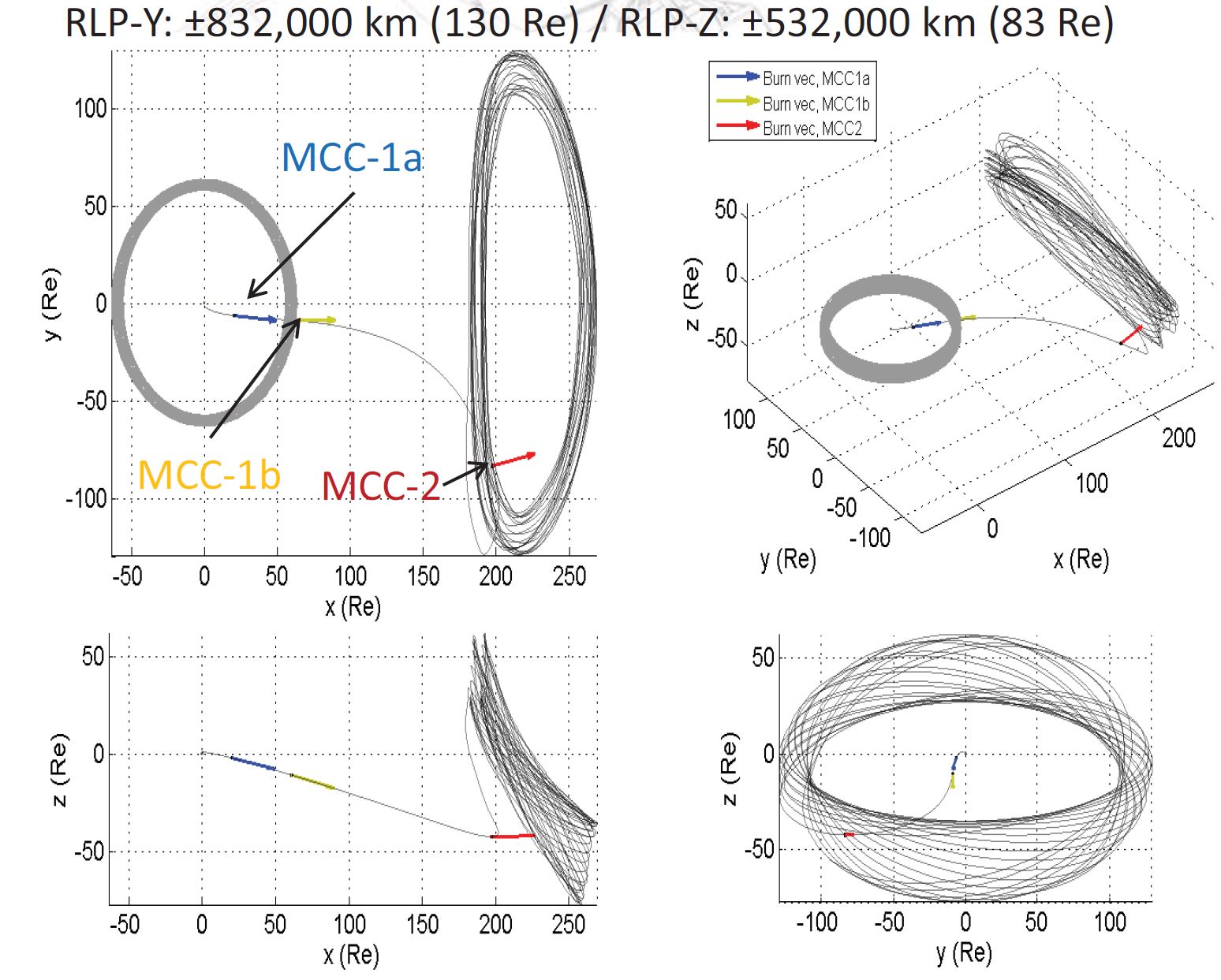}
        \caption{Representation of the deep space transfer orbit from Earth to the L2 operational orbit. The maneouvres are identified as Mid Course Corrections (MCC). Picture from the work of Petersen et al. \cite{petersen2014james}.}
        \label{fig:OrbTransJWST}
    \end{figure}

    Referring to the JWST mission profile, the orbital transfer from Earth to the L2 libration point included three distinct orbital manoeuvres to carefully tune the trajectory and precisely enter the designed operational orbit around L2, as discussed in the paper by Petersen et al. \cite{petersen2014james} and shown in Fig.\ref{fig:OrbTransJWST} extracted by the aforementioned paper.
    This paper reports the results of a Monte Carlo simulation to evaluate the $\Delta V$ to grant the desired orbital path considering the uncertainties related to the engine performance and thrust times. As a result, the manoeuvres are defined in terms of statistical distributions of the thrust times $\Delta T$ and $\Delta V$. The variation of the $\Delta T$ is less effective than that of the $\Delta V$, leading to the decision of using the mean $\Delta T$ and the worst-case scenario $\Delta V$ to compute the accelerations. This introduces a great simplification in the acceleration computation that avoids statistical considerations but still provides meaningful results, shown in Tab.\ref{tab:MCCs}. A 5\% margin over the maximum $\Delta V$ is applied to make the computations even more robust. The actual computation of the accelerations is the simple ratio $a_{MCC}=\frac{\Delta V_{max} 0.05}{\Delta T_{mean}}\frac{1}{9.81}[g]$ between the total $\Delta V$ and the $\Delta T$, that has as reasonable - and likely true - assumption that the manoeuvre is performed at constant engine thrust.
    \newline
    These acceleration values are specific to the JWST mission since its data can be retrieved by literature. The deep orbital manoeuvres for future missions will very likely be different, but the many similarities between JWST and future telescopes like Luvoir lead to the expectation of accelerations in the same order of magnitude for these types of space missions. However, the direction of the forces generated by these accelerations on the shells may change significantly depending on the relative direction of the thrust vector with respect to the shells during manoeuvres. The definitive direction of these forces can be known only once the attitude profile during the deep space transfer is fully defined. 
    \newline
    Finally, the vibration profile to evaluate the dynamic loads is not available in the literature for this phase of the mission. Also, future telescopes like Luvoir may employ systems that decouple the dynamic environment of the spacecraft from that of the telescope, making this evaluation potentially completely useless. However, the vibrations induced by the chemical thrusters - if these are used - are for sure way lower than those generated during launch, causing less issues.

    \begin{table}
    \centering
    \caption{\label{tab:MCCs} Accelerations generated on the JWST during the Mid Course Correction Manoeuvres (MCCs). Values computed starting from \cite{petersen2014james}.}
        \begin{tabular}{c | c | c| c| c}
              & Mean $\Delta V [m/s]$ & Max $\Delta V [m/s]$ & Mean $\Delta T [s]$ & $a_{MCC} [g]$  \\
              \hline \hline
            MCC-1a & 22.279 & 23.4 & 4952.28 & 4.82$\,10^{-4}$ \\
            MCC-1b & 1.967 & 4.5 & 455.68 & 0.001 \\
            MCC-2 & 0.712 & 3.0 & 149.4 & 0.002 \\
        \end{tabular}
    \end{table}
    
\section{Repeated accelerations during the mission lifetime}
    \subsection{Station keeping manoeuvres}
    JWST must perform station keeping (SK) manoeuvres every 21 days in order to maintain its designed halo orbit around L2. The total foreseen $\Delta V$ for the whole operative life of 10.5 years is of 24.88 m/s. This translates in around 2.37 m/s yearly and consequently around 0.14 m/s per SK manoeuvre, according to the work of Dichmann et al.\cite{dichmann2014stationkeeping}
    
    Information about the burning time for this operation can not be found and it has been hypothesized based on the fact that MCC2 can be considered as the first station-keeping operation. A linear assumption based on the acceleration has been carried out according to the following equation, where the assumption of constant thrust during the whole station keeping manoeuvre holds:
    
    \begin{equation*}
       \Delta V_{MCC2}:\Delta T_{MCC2}=\Delta V_{SK}:\Delta T_{SK} \rightarrow 0.712:149.4=0.14:\Delta T_{SK} \rightarrow \Delta T_{SK} = \frac{0.14 * 149.4}{0.712} \approx 30s 
    \end{equation*}
    \begin{equation*}
        a_{SK} = \frac{\Delta V_{SK}}{\Delta T_{SK}}\frac{1}{g*} = \frac{0.14}{30}\frac{1}{9.81} = 4.75 \,10^{-4} g
    \end{equation*}
    
    The direction of this acceleration is variable with the attitude of the spacecraft and of the shell with respect to the thrust direction. Due to the agility and the re-pointing capability of Luvoir, this acceleration may act along all possible directions - tangential to the shell surface or perpendicular to it, with all possible combinations -. The worst case scenario is probably when the acceleration acts completely tangential to the shell, but a comparative analysis shall be performed among the various loading conditions to quantitative find the worst case operative scenario and use it for the constraint system design.
    
    A note shall be done: the computations of the acceleration follow a linear assumption based on MCC2 strategy that may result relatively far from reality, especially on a quantitative level. However, the resulting operative time for a SK single manoeuvre looks realistic. In the case of Luvoir, the value of the needed $\Delta V$ will be probably larger due to its bigger dimensions that directly influence the effect of the solar radiation pressure on the orbit and consequently on the actions to counteract it and on the frequency of the momentum damping manoeuvres. SK thruster can also have higher specific impulse leading to shorter burning time and higher accelerations. However, the order of magnitude of SK accelerations previously computed should be realistic and reliable to preliminary dimension a retaining system. Anyway, the computation using real SK strategy data and associated attitude shall be used for a high-fidelity evaluation of the SK accelerations on the shells. 

    \subsection{Slew manoeuvres}
    Precise values of the attitude accelerations acting on the JWST can be found in the work by Karpenko et al.\cite{karpenko2019agility} related to optimization methods for attitude control and slewing speed. These values are summarised in Tab.\ref{tab:agility}, directly extracted from the previously cited paper and they give a clear idea of the order of magnitude of the applied accelerations and consequent times for slewing. 
    
    \begin{table}[h]
    \centering
    \caption{\label{tab:agility} Agility parameters of the JWST according to different modelling and optimization techniques, data extracted from the work by Karpenko et al. \cite{karpenko2019agility}}
        \begin{tabular}{c | c | c| c}
              Sizing rule & $\alpha_{max} [deg/s^2]$ & $\omega_{max} [deg/s]$ & 90° slew time [min] \\
              \hline \hline
            Two-norm conventional & $0.9 \,10^-{4}$ & 0.037 & 47.0  \\
            Two-norm AIS & $1.08 \,10^-{4}$ & 0.045 & 40.2  \\
            $L_{\infty}$ conventional & $1.14 \,10^-{4}$ & 0.048 & 38.4  \\
            $L_{\infty}$ AIS & $1.45 \,10^-{4}$ & 0.060 & 31.9  \\
        \end{tabular}
    \end{table}
    
\begin{figure}
        \centering
        \includegraphics[scale=1.4]{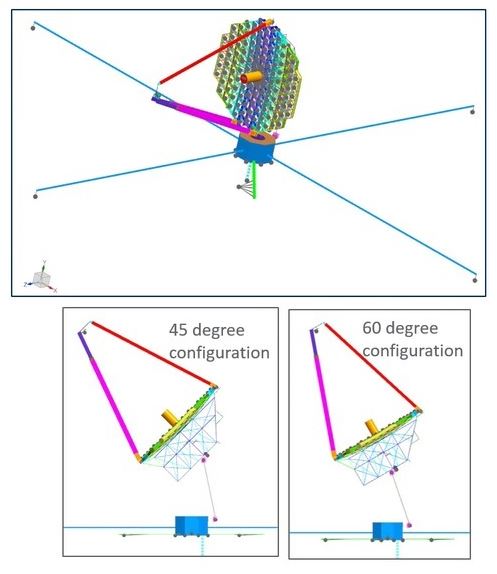}
        \caption{Representation of the gimbal rotation of Luvoir to improve the accessibility time to the complete anti Sun hemisphere. Here the rotation is up to 60° but the maximum capability is of 90°, making the main mirror pointing straight to Nadir. Picture from the work of Tajdaran et al.\cite{tajdaran2018telescope}}
        \label{fig:gimbalLuvoir}
    \end{figure}

\begin{figure}
        \centering
        \includegraphics[scale = 0.7]{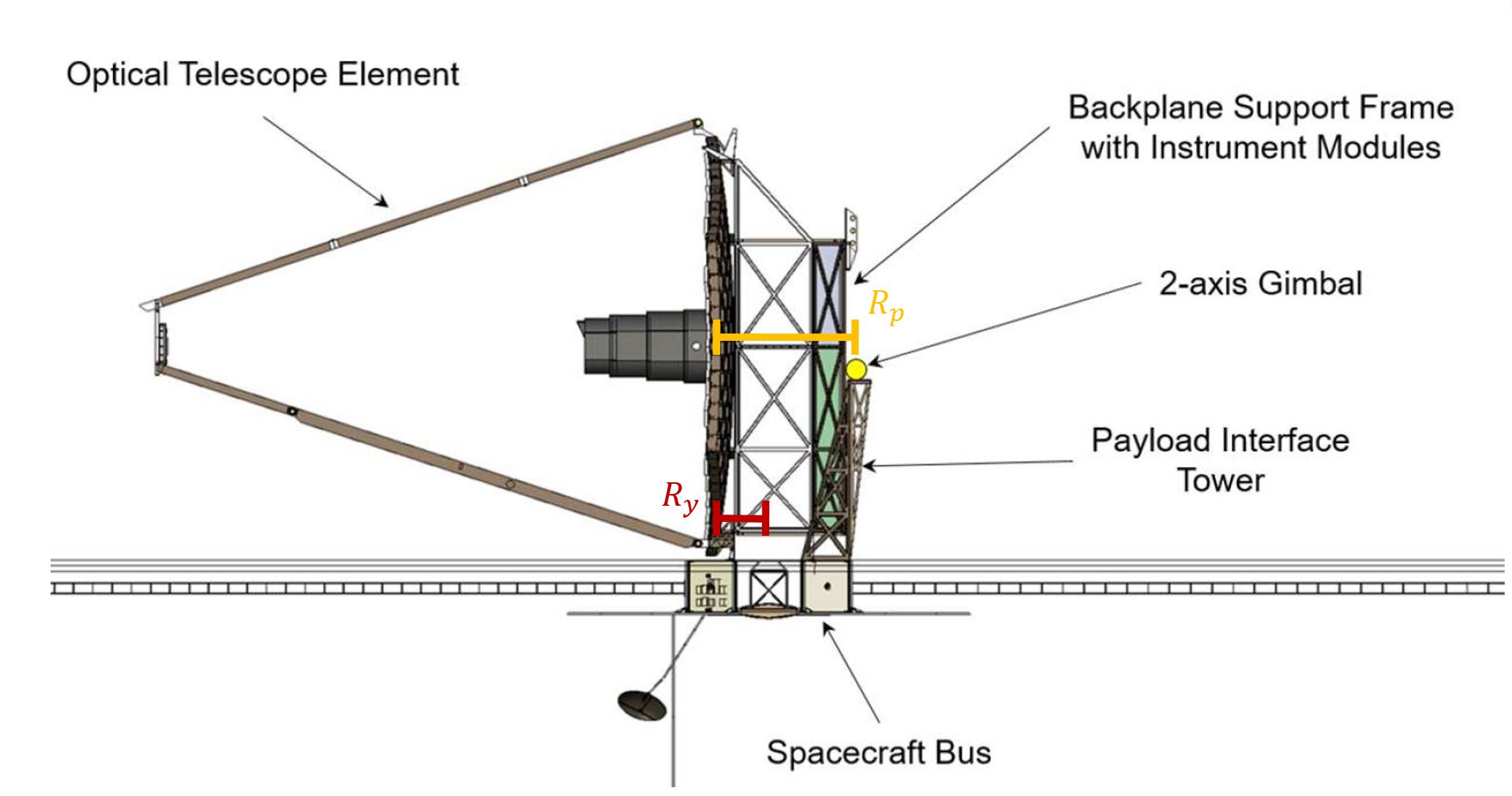}
        \caption{Representation of the distances of the primary mirror from the center of rotations of the Luvoir telescope. Picture elaborated starting from the final design report of Luvoir\cite{luvoir2019luvoir}.}
        \label{fig:distLuvoir}
    \end{figure}

    Literature gives also some interesting information about Luvoir, especially on its strategies to maximise the coverage of the anti-Sun hemisphere. A presentation by Dewell et al.\cite{web:LMLuvoir} places a clear requirement on the re-pointing performance of Luvoir, which shall access any location in the anti-sun hemisphere in a maximum of 45 minutes with 30 minutes as the design goal.  This, together with the geometry of the telescope and especially the distance from the center of rotation, is enough to compute the in-plane accelerations acting on the mirrors that shall be counteracted in order to avoid the misalignment of the magnets and the consequent loss of performance or of the shell. The gimbaling system of Luvoir permits a pitch movement of 90° of the primary mirror and of the tower where the instruments are mounted - discussed in the paper by Tajdaran et al.\cite{tajdaran2018telescope} -, as shown in Fig.\ref{fig:gimbalLuvoir}. However, the gimbal performance and its acceleration cannot be found in literature. The values for this acceleration are nevertheless computed considering the data related to the JWST listed in Tab.\ref{tab:agility} combined with the operative time requirements enlisted in the work by Dewell et al. \cite{web:LMLuvoir} The roll motion is not considered since it would expose the telescope to the sunlight and it would rarely happen and at negligible angular accelerations.
    \newline
    Based on the images of the Luvoir telescope, some dimensions of interest can be retrieved. Also, considering that yaw rotation happens around the central point of the whole spacecraft – a very plausible assumption derived from a symmetric AOCS configuration-, the maximum distance between the primary mirror and the centre of the rotation is  $R_y \approx 2.7m$. Same reasoning for the pitch rotation but considering that it happens at the gimbal point, the distance between the center of rotation and the primary mirror is then $R_p\approx 5.45m$. These distances are highlighted in Fig.\ref{fig:distLuvoir}.
    All the data are now available to evaluate the linear accelerations acting on the mirror due to yaw ($a_y$), pitch ($a_p$) and the combined acceleration ($a_t$) increased by a 10\% margin to compensate for the unknowns related to the pitch rotation and to the AOCS system of Luvoir. This is a very serious worst-case scenario – probably even not feasible due to the limitations of the attitude control system - since all the accelerations are at their maximum at the same time.

    To grant the full coverage of the observable hemisphere, the slew capability shall grant a yaw slew range of 180° and a pitch rotation of 90°, both shall happen in a 30 minutes time window. 
    Based on the assumption of a linear angular motion at a constant acceleration and deceleration - uniformly accelerated angular motion -, some iterations to define the yaw angular acceleration led to the following values:
    \newline
    $\omega_{y,max}=0.135 \, \dfrac{deg}{s}$ is the maximum angular velocity during the yaw rotation, $\alpha_y = 3\,10^{-4} \, \dfrac{deg}{s^2} = 5.2\,10^{-6} \, \dfrac{rad}{s^2}$. Using the uniformly accelerated angular motion assumption,it is possible to compute the acceleration times $t_{acc}$ and the time spent at constant angular speed $t_{\omega}$ to verify that the time constraint of 30 minutes is respected:
    
    \begin{equation*}
       t_{acc} = \frac{\omega_{y,max}}{\alpha_y} \approx 450 s = 7.5 \, mins \rightarrow \theta_{acc} = \frac{1}{2} \alpha_y t_{acc}^2 = 30.375^{\circ}
    \end{equation*}
    \begin{equation*}
        t_{\omega} = \frac{180^{\circ}-2\theta_{acc}}{\omega_{y,max}} \approx 885 s = 15 \, mins \rightarrow t_{180^{\circ}} = t_{\omega} + 2t_{acc} = 30 \, mins \rightarrow \alpha_y,\omega_{y,max} \, OK
    \end{equation*}
    
    As discussed earlier, the pitch rotation of the telescope is performed with the gimbal rotation of the telescope alone. The computations related to its angular acceleration are performed like those of the yaw computation.
    \newline
    $\omega_{p,max}=0.0818 \, \dfrac{deg}{s}$ is the maximum angular velocity during the yaw rotation, $\alpha_p = 2\,10^{-4} \, \dfrac{deg}{s^2} = 3.5\,10^{-6} \, \dfrac{rad}{s^2}$. Again, $t_{acc}$ and $t_{\omega}$ for the pitch rotation can be computed to verify that the time constraint of 30 minutes is respected.
     \begin{equation*}
       t_{acc} = \frac{\omega_{p,max}}{\alpha_p} \approx 414 s = 7 \, mins \rightarrow \theta_{acc} = \frac{1}{2} \alpha_p t_{acc}^2 = 17.14^{\circ}
    \end{equation*}
    \begin{equation*}
        t_{\omega} = \frac{180^{\circ}-2\theta_{acc}}{\omega_{p,max}} \approx 681 s = 11.35 \, mins \rightarrow t_{180^{\circ}} = t_{\omega} + 2t_{acc} \approx 25 \, mins \rightarrow \alpha_p,\omega_{p,max} \, OK
    \end{equation*}
    Finally, the linear accelerations acting on the shells can be computed. The maximum acceleration due to yaw rotation is computed in Eq.(\ref{eq:ay}), the one related to pitch in Eq.(\ref{eq:ap}), where the 10\% margin is added. In Eq.(\ref{eq:at}), the composed acceleration when the shells are subjected to both the yaw and pitch accelerations is computed. 
    \begin{equation}
        a_{y} = 1.1 \,R_y[m]\alpha_y[\frac{rad}{s^2}] = 1.1* 2.7*5.2 \, 10^{-6} \approx 15 \, \mu g
        \label{eq:ay}
    \end{equation}
    \begin{equation}
        a_{p} = 1.1 \, R_p[m]\alpha_p[\frac{rad}{s^2}] = 1.1* 5.45*3.5 \, 10^{-6} \approx 21 \, \mu g
        \label{eq:ap}
    \end{equation}
    \begin{equation}
        a_{t} = \sqrt{a_{p}^2 + a_{y}^2} \approx 26 \, \mu g
        \label{eq:at}
    \end{equation}
    
The computed total acceleration $a_p$ represent the absolute worst case scenario. The related acceleration is the one that shall be counteracted to avoid any sort of misalignment between the shells and their main bodies during the slew manoeuvres. Because of the geometry of the system, these accelerations act on the shell surface and not perpendicular to them, on a first approximation where the primary is considered as flat. The generated forces are shear forces that make the shell slide away from the reference body and a constraining system shall be considered to avoid this effect.
    
\section{Requirements definition for the shell retaining system}
This section places some first level requirements to design the retaining system.They can obviously be further refined or developed, but they place a good solid base. We identified the following:
\begin{enumerate}
    
    \item \textbf{POWER CONSUMPTION}: the retaining system shall not reduce the available power to science instruments while in operations\footnote[2]{The available power is usually designed for the worst case usage scenario adding margin, leaving some power for the retaining system if needed. Alternatively, some power can be dedicated in design phase.}.
    \item \textbf{DYNAMIC BEHAVIOUR}: the retaining system shall not transmit vibrations from the reference body to the shell, apart from low frequency vibrations (< 0.1Hz) and vibrations that induce deformations that are controllable by the actuators\footnote[3]{the retaining system shall not introduce exitations of the intra-actuators modes.}.
    \item \textbf{RIGIDITY}: the retaining system shall limit the translations and rotational displacements of the thin shell within a given bounding box of TBD dimensions\footnote{The dimensions of this box are to be defined according to the assembled system and to the various elements that may enter in contact during the various phases of the lifetime}. The static accelerations that the retaining system shall counteract are listed in Tab.\ref{tab:reqs} for the various expected conditions during lifetime.
    
      \begin{table}[h]
    \centering
    \caption{\label{tab:reqs} Static accelerations affecting the system during the expected phases of the lifetime of a JWST like telescope}
        \begin{tabular}{c | c | c| c}
              \textbf{Life phase} & \textbf{Max acceleration} & \makecell{\textbf{Frequency of acceleration} \\ \textbf{event}} & \textbf{Acceleration direction} \\
              \hline \hline
            Launch & \makecell{$a_{long}\approx4.5g$\\$a_{lat}\approx0.25g$} & Once in the lifetime & \makecell{The main accelerations acts \\ longitudinal wrt flight direction, \\ acting as a shear force that \\ makes the shell slide away from \\the reference body. The lateral\\ accelerations acts both\\ tangentially and perpendicularly\\ to the shell surface} \\
            \hline
            Orbital transfer & $a_{tr}\approx2mg$ & \makecell{Limited number of \\ times in the first months \\ of the mission.\\ JWST  performed 3 of such \\ manoeuvres in the first month} & \makecell{Variable: depends\\ on flight configuration}  \\
            \hline
            Orbital maintenance & $a_{mnt}\approx0.5mg$ & \makecell{Every few days during the\\ whole  operative life.\\ JWST  performed this \\ manoeuvres once every 21 days} & \makecell{Variable: depends\\ on flight configuration. \\ Expected mainly perpendicular \\ to the shell surface}   \\
            \hline
            Slew manoeuvres & $a_{slew}\approx26\mu g$ & Thousands of times & \makecell{Tangential to shell surface. \\ Introduces a shear force} \\
        \end{tabular}
    \end{table}
    
\end{enumerate}

\newpage

\section{Short summary of possible solutions}

As discussed and shown in Tab.\ref{tab:reqs}, the loads that act on the shell/reference body system range from g levels to $\mu g$ levels during the lifetime. The operations happening frequently during the complete operative lifetime do not exceed the $mg$ levels while only during launch the g level accelerations are experienced.
\newline
These levels of accelerations can be counteracted using properly designed mechanical retainers or relaying on the holding capabilities of the magnetic system. 
\newline
Large voice coil mirrors for ground based telescopes are usually designed to sustain static loads up to 1.5g using bias magnets to sustain the weight of the mirror when the actuation magnets are down, as discussed in the works by Riccardi et al. \cite{10.1117/12.858229} about LBT, Briguglio et al. \cite{2014SPIE.9148E..45B} about VLT, Briguglio et al. \cite{briguglio2018optical} about the Magellan telescope and Biasi et al. \cite{2016SPIE.9909E..7YB} about M4 of the E-ELT. This solution can be implemented also in space applications but it shall be carefully tuned and designed to satisfy the much lower loads imposed by operations in space, avoiding useless increases of required power and mass.
Also, the use of diamagnetic materials can be investigated to increase the generated magnetic field at the same power use.
\newline
The launch loads can be counteracted using the electrostatic locking as discussed in Briguglio et al. \cite{briguglio2017toward} and Briguglio et al.\cite{briguglio2018optical}. This strategy requires the use of large amounts of power to feed the coils and to lock the shell to its reference body. It would be possible only using dedicated power sources that like primary batteries on the launcher.

\section{Future developments}

The load analysis can be further refined including the dynamic study of the vibrations environment. Experimental studies are currently carried on to analyse the vibrations damping provided by the voice coil actuation between the reference body and the thin shell. This damping is key to understand the loads that would really affect the shell. Any simulation that does not include this effect may lead to way larger values than the real ones, leading to a not optimal design process for the loads during operations.

\section*{Acknowledgements}
The view expressed herein can in no way be taken
to reflect the official opinion of the European Space Agency. The LATT prototype is property of ESA and has been kindly made available by ESA for laboratory testing with a loan agreement.
The SPLATT project is funded by INAF - Istituto Nazionale di Astrofisica under the TECNO-PRIN INAF 2019 program.
\newpage
\newcommand{\procspie}{Proc. of SPIE}
\bibliography{Biblio.bib}

\bibliographystyle{spiebib}

\end{document}